\documentclass{SCGE}
\usepackage{multicol}
\usepackage{amsmath}
\usepackage{array}
\usepackage{cite}
\begin{document}

\begin{picture}(0,0){\rm
\put(0,-39){\makebox[160truemm][l]{\bf {\sanhao\raisebox{2pt}{.}}
Research Paper  {\sanhao\raisebox{1.5pt}{.}}}}}
\end{picture}

\input psfig.sty
\def\bm{\boldsymbol}

\def\dl{\displaystyle}
\def\du{\end{document}}

\Year{2013} %
\Month{November}
\Vol{56} %
\No{11} %
\BeginPage{2226} %
\EndPage{2235} %
\AuthorMark{{\rm AN Fangxia,} et al.}
\DOI{10.1007/s11433-013-5331-y}%10.1007/s11433-010-0152-8} % The author doesn't need fill in it.

\newcommand{\SIII}{[{S}~{\xiaosanhao {III}}]}
\newcommand{\Msun}{M_{\odot}}
\newcommand{\Ha}{H$\alpha$}
\newcommand{\OIII}{[{O}~{\scriptsize {III}}]}
\newcommand{\NII}{[{N}~{\scriptsize {II}}]}
\newcommand{\FeII}{[{Fe}~{\scriptsize {II}}]}
\newcommand{\OII}{[{O}~{\scriptsize {II}}]}
\newcommand{\SII}{[{S}~{\scriptsize {II}}]}
\newcommand{\SI}{[{S}~{\scriptsize {I}}]}
\newcommand{\HII}{{H}~{\scriptsize {II}}}
\newcommand{\mm}{$\mu$m}
\newcommand\aper[1]{#1"\ diameter}%
\newcommand{\Hb}{H$\beta$}

% \title[short text for running head]{full title}{comments for title}
%\title{The properties of \SIII$\lambda\lambda9096,9532$ Emission-Line Galaxies at $z\sim$1.3}%$^\dagger$}%标题
\title{The nature of \SIII$\lambda\lambda9096,9532$ emitters at $z$ = 1.34 and 1.23}

\author[1,2]{AN FangXia}{}%换手动
\author[2]{ZHENG XianZhong}{}
\author[2]{MENG YanZhi}{}
\author[3]{CHEN Yang}{} 
\author[2]{\\\vspace{3mm}WEN ZhangZheng}{}
\author[1]{L\"{U} GuoLiang}{}%换手动
%\author[1,2,3]{\\\vspace{3mm}LUO YangPing}{}
%\author[1,2*]{HAN ZhanWen}{}

\address[{\rm1}]{School of Physics and Technology, Xinjiang University, Urumqi 830046, Xinjiang, China;}
\address[{\rm2}]{Purple Mountain Observatory, China Academy of Sciences, Nanjing 210008, China;}
\address[{\rm3}]{Center for Astrophysics, University of Science and Technology of China, Hefei 230026, China;}

\maketitle \vspace{-3.5mm}{\footnotesize\begin{center} %Received December 21, 2009; accepted December 30, 2009%收稿日期
\end{center}}\vspace*{-5mm}

\renewcommand{\SIII}{[{S}~{\scriptsize {III}}]}
%     Abstract is required.
\begin{center}
\rule{16.5cm}{0.4pt}
\parbox{16.5cm}
{\begin{abstract} 
A study of \SIII$\lambda\lambda9096,9532$ emitters at $z$ = 1.34 and 1.23 is presented using our deep narrow-band $H_2S1$ (centered at 2.13\,\mm) imaging survey of the Extended Chandra Deep Field South (ECDFS).
We combine our data with multi-wavelength data of ECDFS to build up spectral energy distributions (SEDs) from the $U$ to the $K_{\rm s}$-band for emitter candidates selected with strong excess in $H_2S1 - K_{\rm s}$ and derive photometric redshifts, line luminosities, stellar masses and extinction. 
A sample of 14 \SIII\ emitters are identified with $H_2S1<22.8$ and $K_{\rm s}<24.8$ (AB) over 381 arcmin$^{2}$ area, having \SIII\ line luminosity $L_{\rm [SIII]}= \sim 10^{41.5-42.6}$\,erg\,s$^{-1}$. 
None of the \SIII\ emitters is found to have X-ray counterpart in the deepest Chandra 4\,Ms observation, suggesting that they are unlikely powered by AGN. %an AGN. 
\textit{HST}/ACS F606W and \textit{HST}/WFC3 F160W images show their rest-frame UV and optical morphologies.
About half of the \SIII\ emitters are mergers and at least one third are disk-type galaxies.
Nearly all \SIII\ emitters exhibit a prominent Balmer break in their SEDs, indicating the presence of a significant post-starburst component.
%The modeled SEDs show that the Balmer break and the strong emission lines are the prominent features for nearly all of the \SIII\ emitters.
%This suggests that the presence of post-starburst plus an ongoing burst in the \SIII\ emitters, implying that both shock heating in post-starburst and photoionization caused by young O and B stars are likely to excite the strong \SIII\ emission lines.
Taken together, our results imply that both shock heating in post-starburst and photoionization caused by young massive stars are likely to excite strong \SIII\ emission lines.
We conclude that the \SIII\ emitters in our sample are dominated by star-forming galaxies (SFGs) with stellar mass $8.7<\log (M/\Msun)<9.9$. %extinction $0 < A_{\rm V} <\sim$3\,mag derived from SED modeling and \SIII\ line luminosity $L_{\rm [SIII]}= \sim 10^{41.5-42.6}$\,erg\,s$^{-1}$. 
\end{abstract}}
\end{center}\vspace*{-0.6cm}
\begin{center}
\parbox{16.5cm}{\bf\jiuhao galaxies: evolution, galaxies: high-redshift, galaxies: emission lines, \SIII\  %关键词
}
\end{center}

\begin{center}
\parbox{16.5cm}{\PACS{\hspace*{-2mm}\rm 98.62.Ai, 98.62.Lv, 98.62.Ck}%分类号
\rule{16.5cm}{0.4pt}}\end{center}

%--------------------------------citation---------------------------------------------

%--------------------------------citation---------------------------------------------

%%%%%%%%%%%%%%%%%%%%%%%%%%%%%%%%%%%%%%%%%%%%%%%%%%%%%%%%%%%%
\wuhao\vspace*{1.5mm}
\begin{multicols}{2}
%%%%%%%%%%%%%%%%%%%%%%%%%%%%%%%%%%%%%%%%%%%%%%%%%%%%%%%%%%%%
%% Text of article.
%%%%%%%%%%%%%%%%%%%%%%%%%%%%%%%%%%%%%%%%%%%%%%%%%%%%%%%%%%%%
%    Section headings
\renewcommand{\baselinestretch}{1.08} \baselineskip 12.2pt\parindent=10.8pt

\no %正文

\section{Introduction}
The cosmic star formation rate density peaks at $z\sim2-3$ and sharply declines to the present day\cite{Madau96,Hopkins06,Sobral13}.
The determination of physical properties of galaxies over all cosmic epochs is key to our understanding of physical processes driving the strong evolution in galaxy star formation (SF). 
Galaxies in the ``redshift desert'' 1.2$<z<2.5$ are seldom explored with optical spectroscopies because the major emission lines are redshifted into the near-infrared (NIR), although substantial progress has been made in measuring the luminosity function \cite{Marchesini12}, stellar mass function \cite{Karim11}, dust extinction \cite{Sobral09} and metallicity \cite{Yuan13} of galaxy
populations in this cosmic epoch. Moreover, physical conditions such as ionization state and element abundance can only be derived from important nebular emission lines such as \OIII$\lambda$5007, \NII$\lambda$6584 and \SIII$\lambda$9069.
This can be done with spectroscopic surveys in the NIR with Fiber Multi Object Spectrograph on the Subaru Telescope and with NIR grisms of Wide-Field-Camara 3 (WFC3) on board the Hubble Space Telescope (\textit{HST}), allowing to observe the \NII, \Ha\ emission lines in galaxies at 1.2$<z<$1.5. 
The emission lines at longer wavelengths (for instance, \SIII) are missed in such surveys \cite{Yabe12, Dominguez13}.

\SIII$\lambda\lambda$9096,9532 is widely used to probe the physical conditions of galaxies and nebulae in the local universe \cite{Zastrow11,Garnett89,Perez03,Kehrig06,Matrozis13,Hagele06,Kistiakowsky93}. 
Two major ionization stages are often observed for Sulfur,  S$^{+}$ and S$^{++}$. S$^{++}$ becomes more abundant in high-ionization zone \cite{Kistiakowsky93}, and is believed to be the dominant (often $>$ 60\%) sulfur ion in a nebula \cite{Garnett89}.
S$^{++}$ has three forbidden transitions at \SIII$\lambda\lambda$9096,9532 \,\AA\ and $\lambda$6312\,\AA.
The \SIII$\lambda$6312 line is weak and sensitive to environment temperature, while the \SIII\ lines in the NIR are usually considered much stronger.
These \SIII\ lines are hence often used to estimate the Sulfur abundance for low-$z$ galaxies \cite{Garnett89,Perez03,Kehrig06}.
Sulfur is one of the $\alpha$-elements and its abundance in stellar atmospheres is needed for exploring galaxies formation and evolution \cite{Matrozis13}.
In \HII\ regions, the intensities of \SIII\ emission lines depend exponentially on the electron temperature \cite{Perez03}.
The line ratios between NIR \SIII\ and \SIII$\lambda$6312\,\AA\ are widely used to derive electron density and temperature in \HII\ galaxies \cite{Perez03,Kehrig06,Hagele06}.
In the case of photoionization, \SII\ is mostly yielded below 10 000\,K and \SIII\ increases until 20 000\,K \cite{Kistiakowsky93}.
The abundance of \SIII\ relative to \SII\ is used as an indicator of ionization conditions and the line flux ratios of them are often used to trace the high-excitation for nearby galaxies \cite{Zastrow11}.
%The line flux ratios of \SIII\ to \SII\ are often used to trace the high-excitation for nearby galaxies [10]. 
%Moreover, the abundance of \SIII\ relative to \SII\ is used as an indicator of ionization conditions in the case of photoionization.
Diaz et al. \cite{Diaz85a} pointed out that the \SIII\ lines are weak compared to \SII, \OII\ or \OIII\ lines in the case of collisional ionization because of the shock waves with velocities of $<$ 100\,km s$^{-1}$ \cite{Dopita77}. %S$^{++}$, in shocked gas, cools predominantly via UV line emission, corresponding to higher temperatures, rather than by emission in the NIR [18].
However, the shock models of Binette et al. \cite{Binette85} suggest that the \SIII\ can be enhanced by high-velocity shocks ($>$ 130\,km s$^{-1}$), although \SIII$\lambda\lambda$9096,9532 is still weaker than \SII$\lambda\lambda$6717,31.
More research effort is needed to examine the excitation conditions of \SIII\ in terms of photoionization versus shock heating \cite{Kistiakowsky93,Diaz85a,Diaz85b,Hill99}.

The \SIII\ lines will be widely observed in upcoming NIR spectroscopic surveys.
The NIR narrow-band imaging, a modest way to identify emission-line galaxies with relatively-precise redshift ($\delta z/(1+z)\sim1$\%) over large sky coverage, is widely used to explore high-redshift universe \cite{Sobral09,Ly12} and distinguish environments (for example, groups or clusters) traced by the emission-line galaxies \cite{Tadaki12,Grutzbauch12}.
It is therefore important to ascertain the nature of \SIII\ emitters at high redshifts in order to better understand the yield of these NIR narrow-band surveys.

Herein we present a sample of 14 \SIII$\lambda$9096,9532 emission-line galaxies at $z$=1.34 and 1.23 identified from our NIR narrow-band imaging survey.
A careful analysis of physical properties of \SIII\ emitters is carried out with deep multi-wavelength data in ECDFS.
%, including the unique X-ray 4\,Ms observation from Chandra [25], \textit{HST}/ACS, \textit{HST}/WFC3 imaging data [26,27,28,29], ground-based UV and optical imaging data from the Multiwavelength Survey by Yale-Chile (MUSYC) [30].
We describe our observation and data reduction as well as give a selection of \SIII\ emitters.
Throughout the paper we adopt a cosmology of [$\Omega_{\Lambda}$, $\Omega_M$, $h_{70}$] = [0.7, 0.3, 1.0]. 
Kroupa initial mass function (IMF) \cite{Kroupa01} and AB magnitude system \cite{Oke74} are used unless otherwise stated.

\section{The Data}%\label{s:obs}
\subsection{Observation and data reduction of $H_2S1$-band data}
Our survey in ECDFS ($\alpha$=03:28:45, $\delta$=$-$27:48:00) was carried out with WIRCam on board CFHT through the $H_2S1$ narrow-band filter ($\lambda_\mathrm{c}=2.130$\,\mm, $\Delta \lambda=0.0293$\,\mm) \cite{Puget04}. 
WIRCam consists of four 2048$\times$2048 HAWAII2-RG detectors, providing a field of view of 20$'$$\times$ 20$'$ and a 0.3$''$ pixel scale.
To cover the gaps between detectors and bad pixels, a dithering technique was adopted in our observations.
The $H_2S1$ observations were made in semester 2011B with total integration time 17.24\,hrs under the seeing conditions between 0.6$''$ and 0.8$''$.

The data was reduced using an Interactive Data Language (IDL) based pipeline called \textit{SIMPLE} (Simple Imaging and Mosaicking Pipeline) \cite{Wang10,Hsieh12}.
The pipeline is used for flat-fielding, subtracting background, removing cosmic ray and instrumental features like crosstalk and calibrating the astrometry and photometry.
Because of the rapidly varied sky color in NIR, exposures from the same dithering block (within 40 minutes) and the same detector are processed in a time and stacked into one background-subtracted science image.
%The images in a dithering block and observed by same detector are stacked into one background-subtracted science image.
After that, we mosaic the four background-subtracted science images from different detectors into a frame science image.
%larger science image which covers all the field of view.
The final mosaic $H_2S1$ image has 383 arcmin$^{2}$ area with integrated exposure time $>10$\,hrs.
We limited our source detection in this area, having an effective 5$\sigma$ limiting magnitude of 22.8\,mag.

\subsection{Multi-wavelength data}
The $K_\mathrm{s}$-band ($\lambda_\mathrm{c}=2.146$\,\mm, $\Delta\lambda=0.325$\,\mm) imaging of ECDFS was obtained with CFHT/WIRCam in semesters 2009B and 2010B \cite{Hsieh12}.
The data are reduced and calibrated in the same way as we describe above for the $H_2S1$ data.
The $K_\mathrm{s}$ image reaches a 5$\sigma$ depth of $K_\mathrm{s}$ = 24.8\,mag in the region of $H_2S1$ source detection.
Wang et al. \cite{Wang10} and Hsieh et al. \cite{Hsieh12} provide more details.

The multi-wavelength data we used in this work include the CFHT/WIRCam $J$-band data from Taiwan ECDFS Near-Infrared Survey \cite{Hsieh12}, $U$, $B$, $V$, $R$ and $I$-band data from the Multiwavelength Survey by Yale-Chile (MUSYC) \cite{Gawiser06}, \textit{HST}/ACS F606W ($V_{606}$) and F850LP ($z_{850}$) imaging from the Galaxy Evolution from Morphology and SEDs \cite{Rix04,Caldwell08}, \textit{HST}/WFC3 F125W ($J_{125}$) and F160W ($H_{160}$) imaging from the Cosmic Assembly Near-infrared Deep Extragalactic Legacy Survey (CANDELS) \cite{Grogin11,Koekemoer11}. 
We summarize 12 bands data in Table 1.
\begin{center}
\noindent {\footnotesize{\bf Table 1}\quad Summary of the 12 bands data used in this work}\vspace{2mm}\\
\footnotesize \doublerulesep 0.2pt %\tabcolsep 23pt 
\begin{tabular*}{0.47\textwidth}{llccc}
\hline\hline 
Camera & Filter & $\lambda_\mathrm{c}$\,(\AA) & FWHM\,(arcsec) & 5\,$\sigma$ depth (AB) \\
\hline 
CTIO   & $U$        & 3507                         & 1.05           & 25.9 \\
       & $B$        & 4600                         & 1.01           & 26.5 \\
       & $V$        & 5379                         & 0.94           & 26.7 \\
       & $R$        & 6516                         & 0.83           & 26.4 \\
       & $I$        & 8659                         & 0.96           & 24.3 \\
\hline
ACS    & F606W      & 5958                         & 0.12           & 28.5 \\
       & F850LP     & 9052                         & 0.12           & 27.3 \\
WFC3   & F125W      & 12493                        & 0.14           & 27.2 \\
       & F160W      & 15432                        & 0.15           & 26.7 \\
\hline
WIRCam &$J$         & 12540                        & 0.79           & 25.4 \\
       &$K_{\rm s}$ & 21498                        & 0.75           & 24.8 \\
       &$H_2S1$     & 21301                        & 0.80           & 22.8 \\
\hline
\end{tabular*}
\end{center}

\section{Selection of \SIII\ Emitters} %\label{s:sample}
\subsection{Selection of emission-line objects}
SExtractor \cite{Bertin96} is used to detect sources and measure their fluxes in the $H_2S1$-band image.
Our detection of a source requires a minimum of 5 contiguous pixels above $2.5\sigma$ of the background noise.
Exposure map is used as weight image to reduce spurious detections in low signal-to-noise (S/N) regions.
We utilize the dual-image mode in SExtractor to perform photometry in the $K_\mathrm{s}$-band data.
Total 8720 sources are securely detected with an S/N ratio $>5$ in both $H_2S1$- and $K_\mathrm{s}$-band images.

The emission-line candidates are selected according to the significance of their narrow-band excess, that is, their $K_{\rm s}-H_2S1$ color \cite{Ly12,Tadaki12,Bunker95}.
A secure and significant narrow-band excess is mainly determined by the background noises of the narrow- and broad-bands, $\sigma_{\rm H_\mathrm{2}S1}$ and $\sigma_{K_{\rm s}}$, as follows: 
\begin{equation}
K_{\rm s}-H_\mathrm{2}S1 > \Sigma \sqrt{\sigma_{K_{\rm s}}^2+\sigma_{\rm H_\mathrm{2}S1}^2},
\end{equation}
where the right side term is the combined background noise of the two bands and $\Sigma$ is the significance factor \cite{Bunker95}.
We select emission-line candidates based upon the criterion of $\Sigma>$ 3.
Details can be seen in An et al. (in preparation). 
Moreover, an empirical rest-frame equivalent width (EW) cat of $EW>50$\,\AA\ \cite{Geach08} is applied to eliminate false excess caused by photon noises for bright objects.

There are 140 objects meeting the selection criteria of $\Sigma>$ 3 and $EW>50$\,\AA.
These objects could potentially be emitters of any emission-line between Pa$\alpha$ at $z=0.14$ and Ly$\alpha$ at $z=16.52$.
We derive photometric redshifts (photo-$z$) for the 140 objects.
Among them, 33 are found to have spectroscopic redshift (spec-$z$) from the MUSYC catalog \cite{Cardamone10}.
%Among them, 33 are found to have spectroscopic redshift from the MUSYC catalog [39], giving $\Delta z/(1+z) \sim 0.128$ for photometric redshifts estimated from broad-band SEDs, where $\Delta z =|z_\mathrm{spec}-z_\mathrm{phot}|$.   

\subsection{Identification of \SIII\ emitters}%\label{s:Photometric Redshifts}
We use 12 bands data as listed in Table 1 to measure fluxes and construct SEDs for the selected 140 emission-line objects.
Only 72 of 140 have $J_{125}$ and $H_{160}$ data because CANDELS observations cover the central part of ECDFS, that is, the GOODS-South region.
The aperture-matched photometry of the 12 bands data are given in Table 2 in units of AB magnitude.
We use these matched colors to establish SEDs for 140 sample targets.
%Details of matching photometry are given in An et al. (in preparation). 

The software tool EAZY (Easy and Accurate Redshifts from Yale) \cite{Brammer11} is used to derive photo-$z$ from SEDs.
The $K_\mathrm{s}$ magnitude is taken as Bayesian prior.
From the previous studies, we know that the NIR selected emission-line objects are mostly, if not all, SFGs \cite{Ly12,Lee12,Hayes10}.
A library of templates is chosen particularly suitable for the SFGs, including five templates generated based on the P{\'E}GASE population synthesis models \cite{Fioc97} and calibrated with synthetic photometry from semi-analytic models and one template of young ($t=50$\,Myr) and dusty ($A_\mathrm{V}=2.75$) starburst \cite{Brammer11}.
The combination of these templates is able to provide models spanning a broad range of galaxies ages and the P{\'E}GASE models provide a self-consistent treatment of emission lines \cite{Fioc97}.
%This template set spans a broad range of galaxies ages and the P{\'E}GASE models provide a self-consistent treatment of emission lines \cite{Fioc97}. 
We adopt ``z\_peak'', the peak with the highest integrated probability in redshift probability function, as our photo-$z$.

A total of 17 emitter candidates have the photo-$z\sim$ 1.34 and 1.23, corresponding to \SIII$\lambda\lambda$9096,9532 lines.
The relatively high abundance of \SIII\ was not reported by other surveys \cite{Sobral13,Geach08,Lee12,Hayes10}. 
We show the best-fit SEDs obtained by EAZY in the run of photometric redshifts as well as the observed $U$, N393 ($\lambda_\mathrm{c}=0.393$\,\mm) (Hao et al. in preparation), $V_{606}$ and $K_\mathrm{s}$-band image stamps of 17 \SIII\ emission-line candidates in Figure 1.
We individually inspected the SEDs and the image stamps of 17 \SIII\ emission-line candidates and found that three objects had no detection in the $U$-band while a secure signal is shown in the N393-band. 
The fitted SEDs of these three objects were higher than upper limit of the $U$-band.
The inconsistence between the upper limit of the $U$-band and the chosen model for the three objects is because the upper limits were not taken into account in SED fitting of EAZY. 
This inconsistence suggests that the objects should locate at a higher redshift and drop out in the $U$-band because of Lyman Break \cite{Shapley11}.  
The three objects are labeled as ``U-band dropout'' in Figure 1.
We numbered the other 14 \SIII\ emission-line candidates and found two emitters, S\_9S and S\_9N, having similar color and separated by $\sim$ 1$''$-2$''$, indicating that they are in the same merger system.
We report that only two objects, S\_5 and S\_8, are \SIII$\lambda$9532 emitters (photo-$z \sim$ 1.23) and the S\_5 is confirmed by spec-$z$.
Finally, we have identified 12 \SIII$\lambda$9096 (two in one merger system) and two \SIII$\lambda$9532 emitters at $z=1.34$ and $z=1.23$, respectively (refer to Table 3). 
\end{multicols}
\begin{center} %\tabcolsep 23pt
\noindent {\footnotesize{\bf Table 2}\quad Photometry of \SIII\ emitters}\vspace{2mm}\\
\footnotesize \doublerulesep 0.2pt
\tiny \begin{tabular*}{0.92\textwidth}{lcccccccccccccc}
\hline
\hline
ID & Mag\_$U$ & Mag\_$B$ & Mag\_$V$ & Mag\_$V_\mathrm{606}$ & Mag\_$R$ & Mag\_$I$ & Mag\_$z_\mathrm{850}$  & Mag\_$J_\mathrm{125}$ & Mag\_J & Mag\_$H_\mathrm{160}$ & Mag\_$K_\mathrm{s}$ & Mag\_$H_2S1$\\
\hline
  S\_1      &25.68 $\pm$0.11    &25.15 $\pm$0.04    &25.07 $\pm$0.05    &25.08 $\pm$0.04    &25.09 $\pm$0.05    &24.55 $\pm$0.18    &24.49 $\pm$0.07   &$\cdots^{c)}$            &23.82 $\pm$0.06   &$\cdots$            &23.59 $\pm$0.08    &22.65 $\pm$0.10  \\
  S\_2      &24.84 $\pm$0.05    &24.61 $\pm$0.03    &24.54 $\pm$0.03    &24.44 $\pm$0.02    &24.34 $\pm$0.02    &23.84 $\pm$0.10    &23.73 $\pm$0.04   &$\cdots$            &23.49 $\pm$0.04   &$\cdots$            &23.42 $\pm$0.06    &22.48 $\pm$0.09  \\
  S\_3     &-99.99$^{a)}$       &26.07 $\pm$0.10    &25.67 $\pm$0.09    &25.62 $\pm$0.05    &25.57 $\pm$0.08   &-99.99              &25.04 $\pm$0.06    &24.18 $\pm$0.02    &24.31 $\pm$0.07    &24.10 $\pm$0.02    &24.01 $\pm$0.09    &22.94 $\pm$0.13  \\
  S\_4      &24.70 $\pm$0.08    &24.18 $\pm$0.02    &24.13 $\pm$0.02    &24.05 $\pm$0.01    &23.97 $\pm$0.02    &23.61 $\pm$0.08    &23.55 $\pm$0.01    &23.06 $\pm$0.01    &23.16 $\pm$0.03    &22.86 $\pm$0.01    &23.01 $\pm$0.04    &22.42 $\pm$0.09  \\
  S\_5      &24.28 $\pm$0.08    &23.69 $\pm$0.02    &23.64 $\pm$0.02    &23.57 $\pm$0.01    &23.49 $\pm$0.02    &22.85 $\pm$0.05    &22.46 $\pm$0.01    &22.22 $\pm$0.00    &22.43 $\pm$0.02    &21.92 $\pm$0.00    &22.07 $\pm$0.02    &21.62 $\pm$0.05  \\
  S\_6     &-99.99              &26.39 $\pm$0.17    &26.07 $\pm$0.17    &25.99 $\pm$0.07    &25.92 $\pm$0.14    &24.75 $\pm$0.28    &24.88 $\pm$0.05    &23.80 $\pm$0.02    &23.82 $\pm$0.06    &23.13 $\pm$0.01    &22.35 $\pm$0.03    &21.71 $\pm$0.06  \\
  S\_7      &25.26 $\pm$0.14    &24.63 $\pm$0.03    &24.53 $\pm$0.03    &24.49 $\pm$0.02    &24.45 $\pm$0.03    &24.02 $\pm$0.11    &23.75 $\pm$0.04   &$\cdots$            &23.27 $\pm$0.04   &$\cdots$            &23.44 $\pm$0.07    &22.79 $\pm$0.13  \\
  S\_8     &-99.99              &-99.99             &-99.99             &25.91 $\pm$0.04   &-99.99              &24.78 $\pm$0.20    &24.16 $\pm$0.02    &23.57 $\pm$0.01    &23.59 $\pm$0.04    &23.32 $\pm$0.01    &22.89 $\pm$0.04    &22.49 $\pm$0.10  \\
 S\_9S$^{b)}$    &$\cdots$            &$\cdots$           &$\cdots$           &23.47 $\pm$0.04   &$\cdots$           &$\cdots$            &22.94 $\pm$0.08    &22.73 $\pm$0.02    &22.73 $\pm$0.02    &22.55 $\pm$0.02    &22.31 $\pm$0.03    &21.75 $\pm$0.06  \\
 S\_9N$^{b)}$    &$\cdots$            &$\cdots$           &$\cdots$           &24.03 $\pm$0.06   &$\cdots$           &$\cdots$            &23.43 $\pm$0.11    &23.34 $\pm$0.03    &23.34 $\pm$0.03    &23.13 $\pm$0.03    &22.80 $\pm$0.03    &22.31 $\pm$0.07  \\
 S\_10      &25.65 $\pm$0.10    &25.54 $\pm$0.06    &25.49 $\pm$0.07    &25.27 $\pm$0.04    &25.06 $\pm$0.05    &24.98 $\pm$0.25    &24.75 $\pm$0.08   &$\cdots$            &24.33 $\pm$0.07   &$\cdots$            &24.18 $\pm$0.10    &23.02 $\pm$0.12  \\
 S\_11      &26.36 $\pm$0.17    &26.01 $\pm$0.08    &25.90 $\pm$0.10    &25.83 $\pm$0.07    &25.77 $\pm$0.09    &25.08 $\pm$0.26    &24.56 $\pm$0.07   &$\cdots$            &24.02 $\pm$0.06   &$\cdots$            &23.69 $\pm$0.06    &22.94 $\pm$0.12  \\
 S\_12      &24.14 $\pm$0.04    &23.73 $\pm$0.02    &23.70 $\pm$0.02    &23.66 $\pm$0.01    &23.62 $\pm$0.02    &23.24 $\pm$0.08    &22.94 $\pm$0.02   &$\cdots$            &22.80 $\pm$0.03   &$\cdots$            &22.55 $\pm$0.03    &22.18 $\pm$0.09  \\
 S\_13$^{b)}$     &$\cdots$            &$\cdots$           &$\cdots$           &24.48 $\pm$0.02   &$\cdots$           &$\cdots$            &23.94 $\pm$0.04   &$\cdots$            &23.28 $\pm$0.05   &$\cdots$            &23.40 $\pm$0.06    &22.72 $\pm$0.13  \\
\hline      
\end{tabular*}
{ \\ \hspace{8mm} a) no detection; b) not resolved in MUSYC observation; c) no observation. \hfill\mbox{}}
\end{center}

\begin{center}
\centerline{\psfig{figure=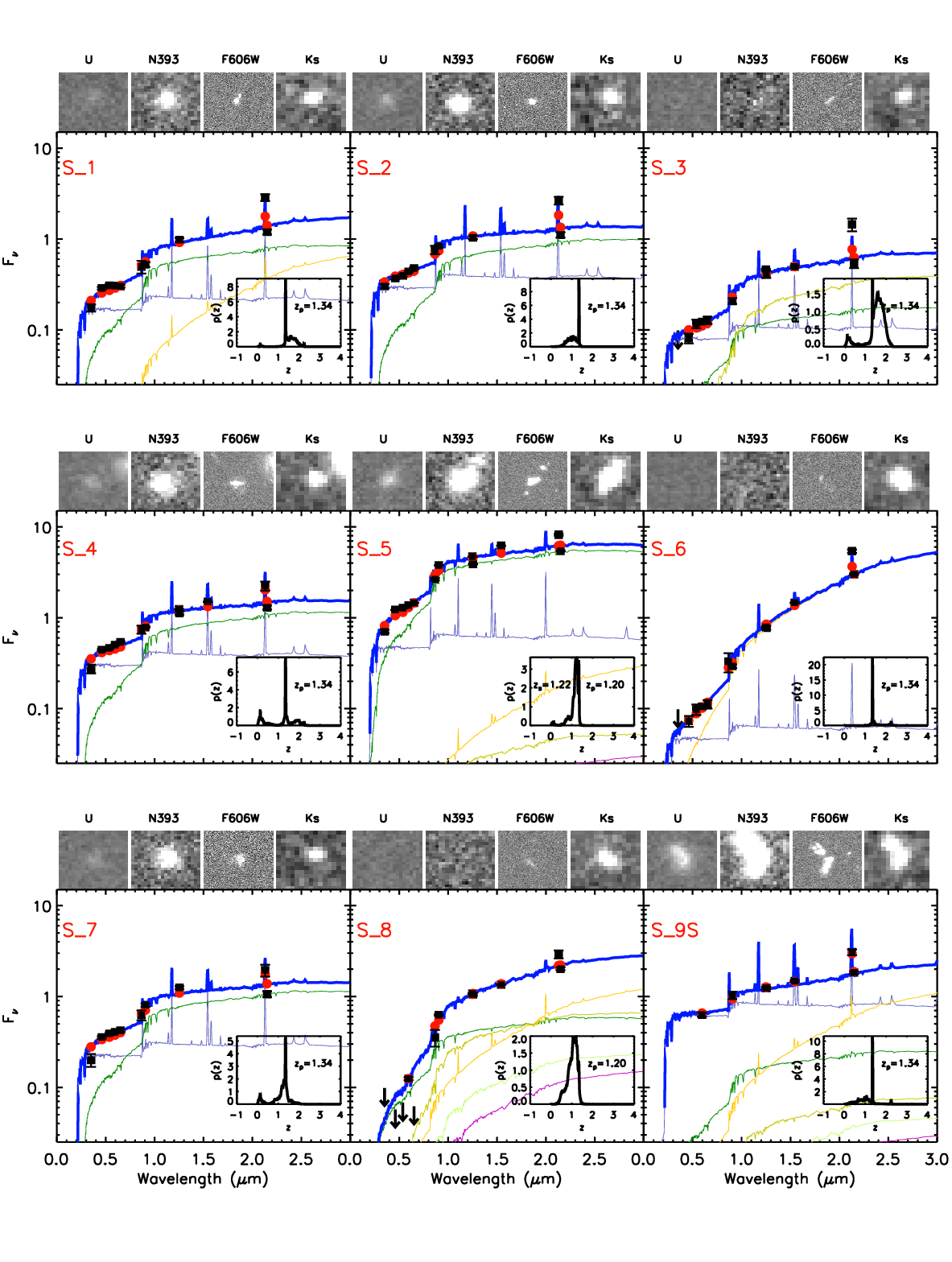,height=22cm}}
\label{f:sed.eps}
\end{center}
\vspace*{-5mm} 
{\footnotesize {\bf Figure.1}\quad Image stamps and SEDs of \SIII\ emitters. Each stamp has a size of $4.5'' \times 4.5''$.
The squares are the observed data points and circles are the best-fit results from EAZY \cite{Brammer11}.
The arrows show the upper limits of corresponded bands which means the source is not detected or resolved in these bands. 
The three U-band dropout objects are marked with a label of ``U-band dropout ''.
The thick line shows the best-fit SED and light lines are the templates.
The inset plot shows the probability distribution of derived photometric redshift.}
%\clearpage

%\vspace*{2mm}
\begin{center}
\centerline{\psfig{figure=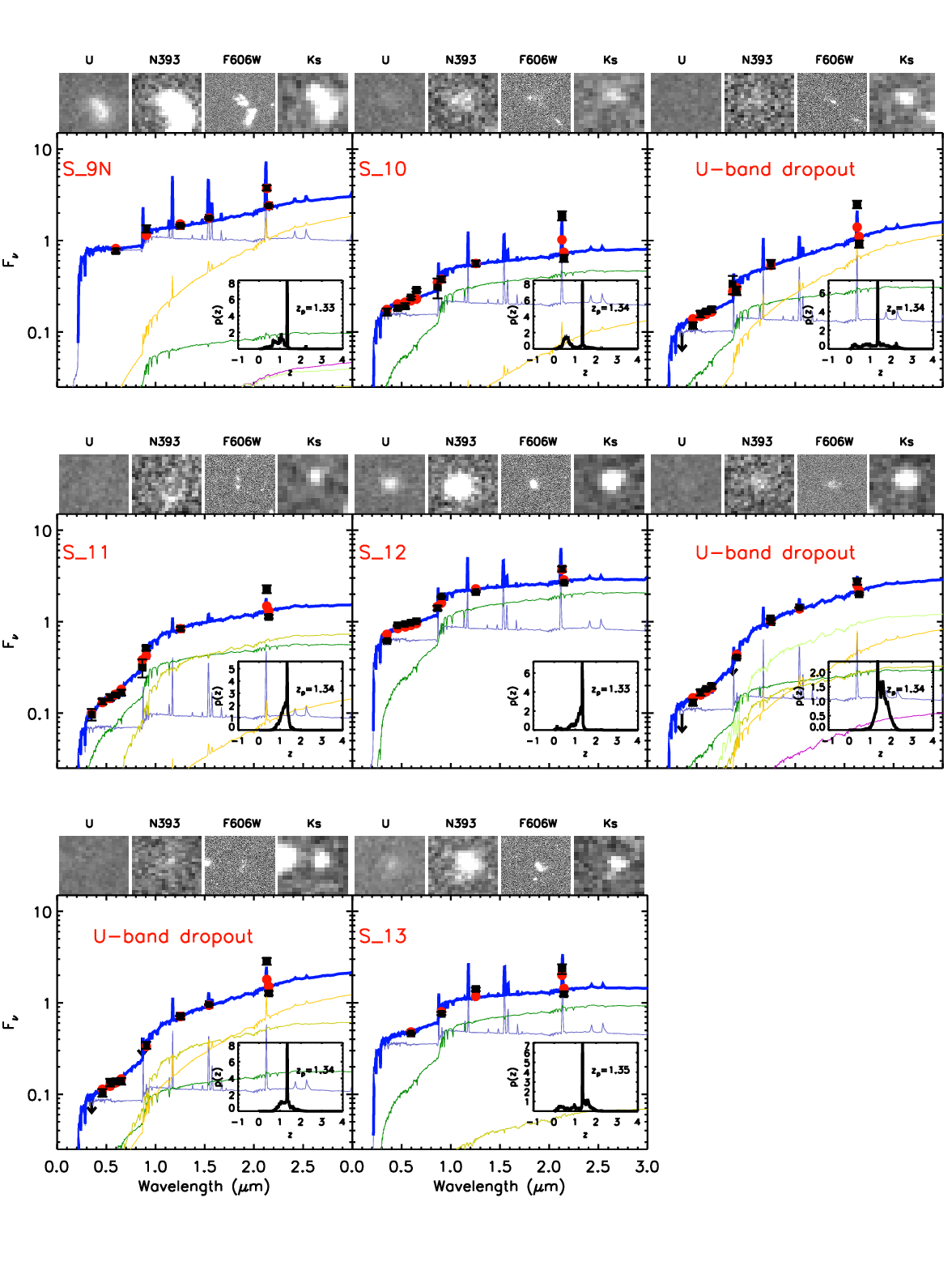,height=22cm}}
\end{center}
\vspace*{-5mm}
{\footnotesize {\bf Figure.1}\quad (continued).}
\clearpage

\begin{center} %\tabcolsep 23pt
\noindent {\footnotesize{\bf Table 3}\quad Properties of \SIII\ emitters}\vspace{2mm}\\

\doublerulesep 0.2pt
\scriptsize\begin{tabular*}{0.96\textwidth}{lcccccccccc}
\hline
\hline
ID &RA  &DEC  & log~L$_\mathrm{[SIII]}$~(erg~s$^{-1}$)  & err\_log~L$_\mathrm{[SIII]}$~(erg~s$^{-1}$) & EW$_\mathrm{obs}$~(\AA) & A$_\mathrm{v}$~(mag)  & log~$(M/\Msun)$  & Photo-$z$ & Spec-$z$ & Morphology\\
\hline
    S\_1  &52.939262  &-27.973326   &41.89     &0.94   &637.84     &0.91     &9.16      &1.35     & $\cdots$   &Clumpy     \\
    S\_2  &53.239098  &-27.971136   &41.76     &0.97   &630.31     &0.00     &9.15      &1.34     & $\cdots$   &Merger    \\
    S\_3  &53.096191  &-27.916807   &41.63     &0.84   &802.79     &0.09     &9.19      &1.35     & $\cdots$   &Disk      \\
    S\_4  &53.222755  &-27.859795   &41.67     &0.87   &318.84     &0.00     &9.63      &1.35     & $\cdots$   &Disk      \\
    S\_5  &53.056183  &-27.855364   &41.85     &0.97   &185.48     &0.11     &9.69      &1.20     &1.226       &Merger    \\
    S\_6  &53.186699  &-27.831648   &42.63     &1.17   &443.30     &2.67     &9.45      &1.35     & $\cdots$   &Disk      \\
    S\_7  &52.909500  &-27.792284   &41.55     &0.69   &365.85     &0.00     &9.24      &1.35     & $\cdots$   &Clumpy   \\
    S\_8  &53.079136  &-27.788351   &41.77     &0.63   &190.99     &1.20     &9.91      &1.20     & $\cdots$   &Disk      \\
   S\_9S  &53.062340  &-27.765062   &42.04     &0.97   &267.93     &0.77     &9.44      &1.34     & $\cdots$   &Merger    \\
   S\_9N  &53.062531  &-27.764765   &41.88     &1.00   &333.47     &0.50     &8.92      &1.34     & $\cdots$   &Merger    \\
   S\_10  &52.967308  &-27.754555   &41.70     &0.89   &998.66     &0.40     &8.67      &1.34     & $\cdots$   &Diffuse   \\
   S\_11  &52.993256  &-27.722013   &41.59     &0.77   &452.96     &0.29     &9.22      &1.34     & $\cdots$   &Diffuse   \\
   S\_12  &52.938370  &-27.719694   &41.72     &0.82   &229.74     &0.00     &9.46      &1.34     & $\cdots$   &Merger    \\
   S\_13  &53.076378  &-27.663435   &41.58     &0.73   &383.50     &0.00     &9.18      &1.35     & $\cdots$   &Merger    \\
\hline
\label{properties}
\end{tabular*}
\end{center}
\wuhao\vspace*{1.5mm}
\begin{multicols}{2}

\renewcommand{\baselinestretch}{1.08} \baselineskip 12.2pt\parindent=10.8pt

\no 

\section{Properties of \SIII\ Emitters}\label{s:property}
We use the multi-wavelength data of ECDFS and modeled SEDs of our sample to analyze the SF activity, X-ray property, morphology, dust attenuation, stellar mass and stellar population of selected \SIII\ emitters.
The derived properties, extinction corrected luminosity, EW, stellar mass, A$_\mathrm{v}$ and morphology type, of our sample galaxies are summarized in Table 3. 
\begin{center}
\centerline{\psfig{figure=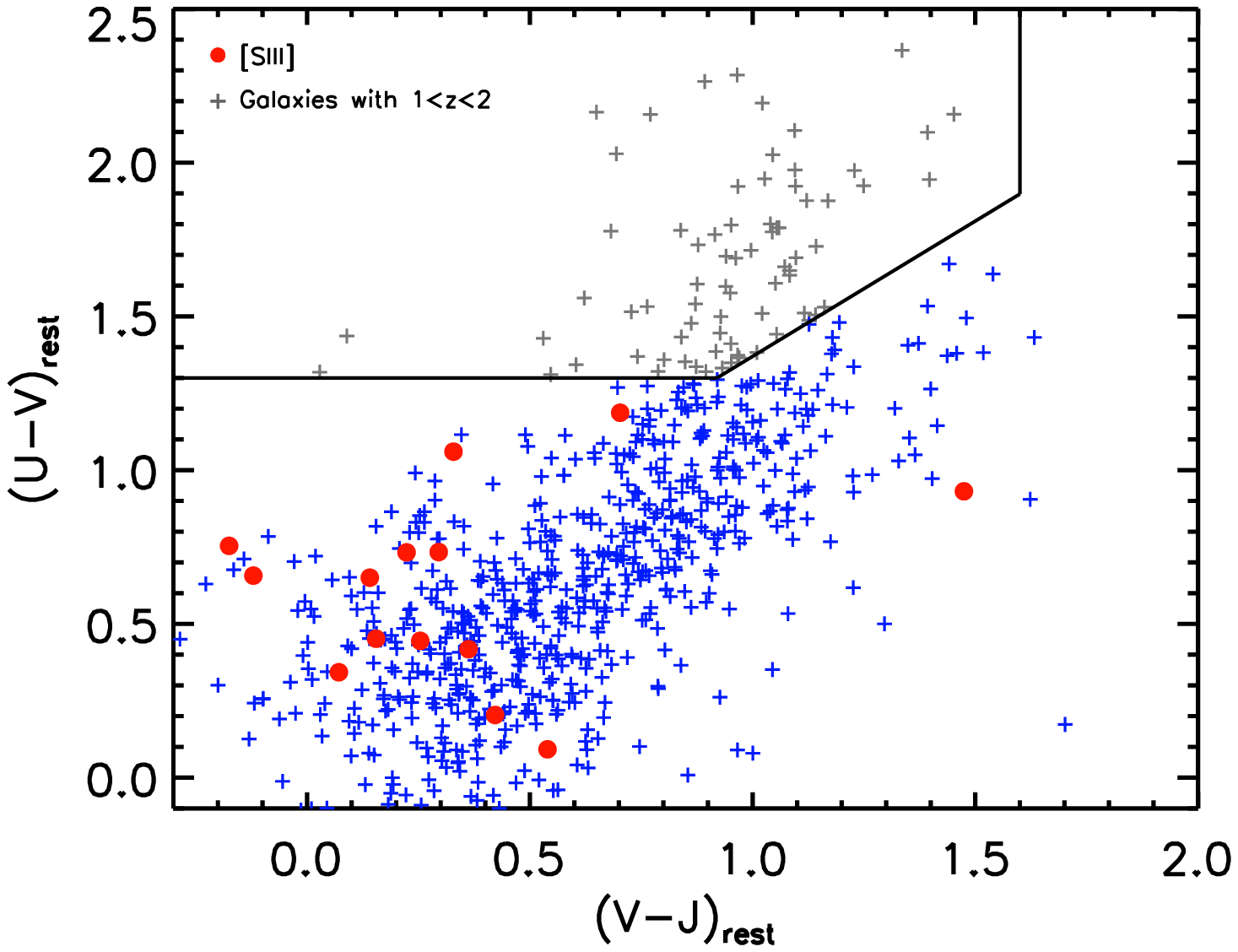,height=7.0cm}}
\label{f:fig3}
\end{center}
\vspace*{-5mm}
{\footnotesize {\bf Figure.2}\quad $U - V$ versus $V - J$ diagram of our \SIII\ emission-line galaxies (red solid points) and $K_\mathrm{s}$-selected galaxies with 1.0 $<z<$ 2.0 from FIREWORKS \cite{Wuyts08}.
The three lines ($V - J >$ 1.6, $U - V >$ 0.49 + 0.88 $\times$ ($V - J$) and $U - V <$ 1.3) define the separation cuts between quiescent galaxies (gray pluses in the top-left region) and SFGs (blue pluses) \cite{Williams09}.}
\vspace*{2mm}
\subsection{Star formation activity and X-ray properties} \label{s:xray}
We adopt a rest-frame color-selection technique \cite{Williams09} to study the SF activity in our sample galaxies.
We employed the criteria of $V - J >$ 1.6, $U - V >$ 0.49 + 0.88 $\times$ ($V - J$) and $U - V <$ 1.3 (three lines in Figure 2) to separate quiescent galaxies from SFGs, following Williams et al. \cite{Williams09}.
We use observed $I$ ($z_{850}$ instead if $I$ is not available), $J$ and $K_\mathrm{s}$ as the rest-frame $U$, $V$ and $J$.
We show our \SIII\ emitters and the $K_\mathrm{s}$-selected galaxies with 1.0 $<z_\mathrm{spec}<$ 2.0 from FIREWORKS \cite{Wuyts08} in Figure 2.
The blue pluses in Figure 2 represent the SFGs from FIREWORKS.
All of the \SIII\ emitters (red solid points) distribute in the locus of these blue SFGs. 

The deepest X-ray data from Chandra 4\,Ms observation allow the X-ray properties of our sample galaxies to be examined \cite{Xue11}.
None of the \SIII\ emitters is found to have X-ray counterpart suggestive of no presence of Active Galactic Nuclei (AGN).

\subsection{Morphologies} \label{s:morphology}
We use \textit{HST}/ACS F606W imaging data to examine morphologies of our sample of 14 \SIII\ emission-line galaxies.
This band observe the rest-frame 2546\,\AA\ and 2671\,\AA\ for $z$ = 1.34 and $z$ = 1.23, respectively. 
The \textit{HST}/WFC3 F160W imaging data from CANDELS show rest-frame optical morphologies for half of the sample galaxies.
%corresponding to rest-frame optical band, imaging data from CANDELS are available for half of the sample galaxies.
Here we execute a visual classification of the rest-frame UV morphologies for 14 \SIII\ emission-line galaxies (by An F.X, Zheng X.Z and Wen Z.Z).
We found that 43\% (6/14) sample galaxies appear to be merger remnants with tail, double core; 29\% (4/14) are diffuse; 14\% (2/14) show a readily apparent disk and the remain 14\% have clumpy morphologies.
We compared the rest-frame optical morphologies from CANDELS of seven \SIII\ emission-line galaxies in GOODS-South and found two are different from rest-frame UV morphologies.
We present the F606W and F160W image stamps of these seven \SIII\ emission-line galaxies in Figure 3, they are S\_3, S\_4, S\_5, S\_6, S\_8, S\_9S and S\_9N from left to right, one can see the apparent discrepancy between rest-frame UV and rest-frame optical morphologies of S\_5 and S\_6.
We argue that this discrepancy is caused by the heavily dust extinction for rest-frame UV light which is confirmed by our estimation of dust extinction as described in next subsection (refer to Table 3).
Because of this heavily dust extinction, the S\_5 and S\_6 appear diffuse morphologies in rest-frame UV while they are disk galaxies as shown in rest-frame optical images.
We inspected our morphological classification with F160W images for the other \SIII\ emission-line galaxies in the GOODS-South.
The S\_5 and the S\_6 are modified to disk morphology and the classification of other five \SIII\ emitters are confirmed by rest-frame optical morphologies.
Therefore, our classification based on rest-frame UV morphologies has large uncertainty caused by dust extinction.
Moreover, the morphological parameters, Gini and M$_\mathrm{20}$ coefficient \cite{Lotz04} which are derived from rest-frame UV images, can not delineate the difference between mergers and non-mergers of our sample galaxies as Stott et al. \cite{Stott13} reported.
%\vspace*{2mm}
\begin{center}
\centerline{\psfig{figure=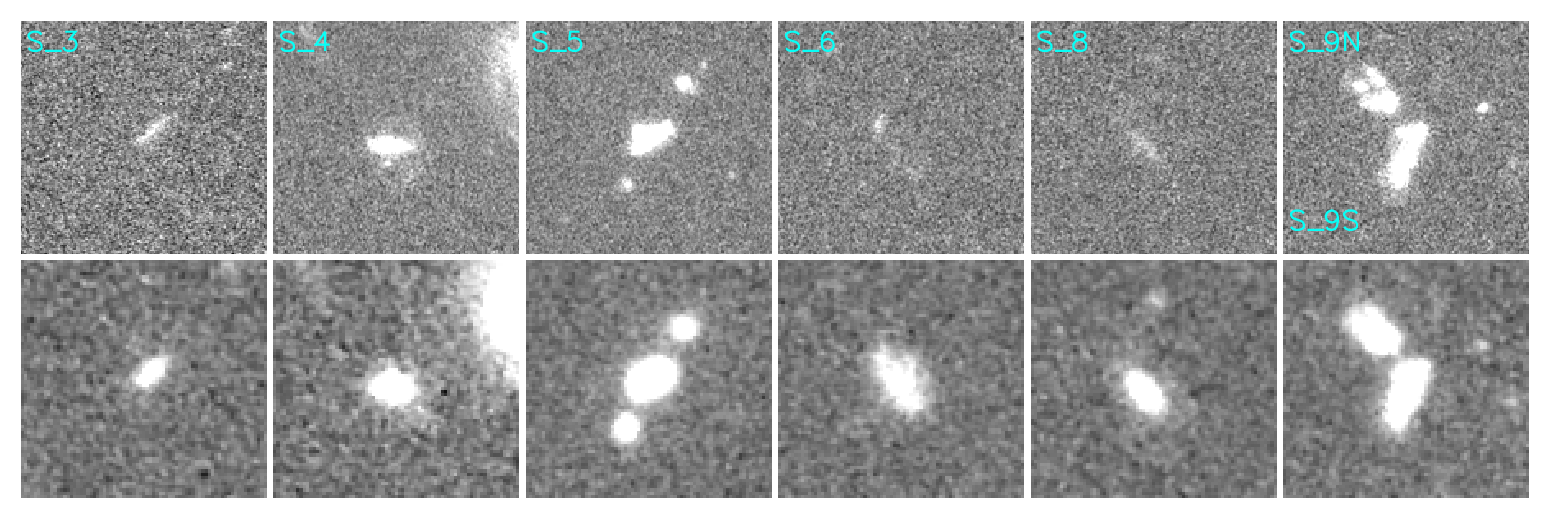,height=3.0cm}}
\label{f:fig2}
\end{center}
\vspace*{-5mm}
{\footnotesize {\bf Figure.3}\quad \textit{HST} F606W (upper) and F160W (bottom) image stamps of seven \SIII\ emission-line galaxies at $z\sim$ 1.3. Each stamp has a size of $4.5'' \times 4.5''$, corresponding to 37\,kpc$\times$37\,kpc at $z\sim$ 1.3.}
\vspace*{2mm}

\subsection{Dust extinction}%\label{s:extinction} 
We estimate dust extinction of \SIII\ emission-line galaxies from the best-fit SEDs as shown in Figure 1.
As described above, we made use of EAZY \cite{Brammer11} to fit SEDs with a library of six galaxy templates.
One of the six templates represents a young and dusty starburst with $A_\mathrm{V}=2.75$ which is presented by yellow light line in Figure 1 (refer to color version in on-line journal).
Other five templates represent stellar populations of different ages and their synthetic photometry are calibrated by semi-analytic models which are unable to produce extremely dusty galaxies \cite{Brammer11,Fioc97}.
Generally speaking, since stars are formed in dusty environments, the young stellar population being most affected by dust extinction.
We assume that the young population comprises two parts: unattenuated and attenuated.
Thus, in this model the amount of dust attenuation in the modeling of a galaxy SED can be estimated through the fraction of the dusty starburst template to the total (refer to Table 4).
We calculated $A_\mathrm{V}$ for \SIII\ emission-line galaxies in our sample using the best-fit templates obtained in the run of photometric redshifts as shown in Figure 1. 
The maximal $A_\mathrm{V}$ in this model is limited by $A_\mathrm{V}=2.75$ adopted in the young and dusty starburst template.
From our SED modeling we find that this limitation has little influence to our extinction estimation of \SIII\ emitters except one, the S\_6, whose SED is dominated by this dusty starburst template (see Figure 1 and Table 4).
We list the estimated $A_\mathrm{V}$ of our sample in Table 3 and we do not find the correlation between our SED-derived extinction and luminosity, stellar mass or line equivalent width of \SIII\ emitters as other emission-line selected samples at similar redshift found \cite{Sobral09,Dominguez13}.

\subsection{Stellar masses} 
We use the software FAST (Fitting and Assessment of Synthetic Templates) \cite{Kriek09} to fit MA05 stellar population synthesis models \cite{Maraston05} with an exponentially declining Star formation histories (SFHs).
Our choice of SFHs guarantees consistency with most current researches \cite{Wuyts11}.
The metallicity is fixed to solar (Z = 0.02) when fit MA05 models.
The SF timescale, $\tau$, varies between 10\,Myr and 10\,Gyr in steps of 0.5\,dex and the age is allowed to vary between 0.1\,Myr and the age of the universe at the observed redshift in estimating stellar mass for our sample.
Dust extinction is modeled using the Calzetti reddening law \cite{Calzetti00}, with A$_\mathrm{v}$ ranges between 0 and 3\,mag in steps of 0.01\,mag.
%We derive model SED for a grid with $\tau$ between 10\,Myr and 10\,Gyr in steps of 0.5\,dex and age in steps of 0.2\,dex with a minimum age of 0.1\,Myr and the maximum age not exceeding the age of the universe.
Photometric redshift derived by EAZY is used as recommended by FAST \cite{Kriek09}.
Although, the two codes use different template sets, FAST can deal with this difference and obtain a reliable result of stellar mass, refer to the appendix in Kriek et al. \cite{Kriek09}.
The estimated stellar masses of our sample galaxies are $8.7<\log (M/\Msun)<9.9$.
\begin{center} %\tabcolsep 23pt
\noindent {\footnotesize{\bf Table 4}\quad Percentages of six best-fit templates at $\lambda$ = 2.146\,\mm}\vspace{2mm}\\
\doublerulesep 0.2pt
\scriptsize\begin{tabular*}{0.47\textwidth}{lcccccc}
\hline
\hline
ID &Temp\_1$^{a)}$ &Temp\_2 &Temp\_3 &Temp\_4 &Temp\_5 &Temp\_6$^{b)}$\\
\hline
    S\_1   &0    &16     &56     &1      &0     &27 \\
    S\_2   &0    &30     &70     &0      &0      &0 \\
    S\_3   &0    &17     &26    &53      &2      &2 \\
    S\_4   &0    &27     &73     &0      &0      &0 \\
    S\_5   &0    &10     &86     &1      &0      &3 \\
    S\_6   &0     &2      &0     &0      &0     &98 \\
    S\_7   &0    &22     &78     &0      &0      &0 \\
    S\_8   &5     &0     &26    &27      &9     &33 \\
   S\_9S   &1    &45     &13     &4      &2     &35 \\
   S\_9N   &1    &46      &4     &1      &1     &47 \\
   S\_10   &0    &28     &60     &1      &0     &11 \\
   S\_11   &0     &7     &40    &47      &0      &6 \\
   S\_12   &0    &30     &70     &0      &0      &0 \\
   S\_13   &0    &34     &62     &4      &0      &0 \\
\hline
\label{temp}
\end{tabular*}
{ \\  a) Temp\_1 to Temp\_5 are five templates generated based on the P{\'E}GASE models;
  \\ \hspace{2mm} b) Temp\_6 is the template of young and dusty starburst. \hfill\mbox{}}
\end{center}
\subsection{Stellar populations} 
The best-fit EAZY templates, which are shown by light color lines in Figure 1, which represent the fractions of different stellar components of modeled SEDs, thus, can be used to analysis the stellar populations of our sample galaxies.
We present the fractions of six templates at $\lambda$ = 2.146\,\mm\ for 14 modeled SEDs in Table 4.
One can see that the template Temp\_3, which is shown by green line in Figure 1 (refer to color version in on-line journal), is the dominant for most of modeled SEDs.
From Figure 1, we see that this template has a significant Balmer break at $\sim$ 3650\,\AA\ (redshifted to $\sim$ 8540\,\AA\ at $z =$ 1.34).
%The Balmer break is the typical spectra shape of A-star represents the stellar component of age = 100\,Myr [40,41].
The Balmer break is the typical spectra feature of A-star which represents the stellar populations with ongoing SF over sustained timescales ($>$ 100\,Myr) or the post-starburst populations 0.3-1\,Gyr since the cessation of SF \cite{Shapley11}.
For the majority of modeled SEDs, the secondary contributor is the template Temp\_2, which is shown by light blue line in Figure 1.
%This template represents the stellar population of young and ongoing SF with blue continuum and strong emission lines. 
This template represents a stellar component of age around 10\,Myr with blue continuum and strong emission lines \cite{Kennicutt12}.
Moreover, the contribution from the dusty, starburst template can not be ignored for most of modeled SEDs and we see that one of modeled SEDs (S\_6) is dominated by this template (refer to Figure 1 and Table 4).
Despite this, from Figure 1, we see that the Balmer break and strong emission lines are the prominent features of SEDs for nearly all of the \SIII\ emitters.
%This indicates that a single stellar component of age $>$ 100\,Myr in combination with a secondary component of age $\sim$ 10\,Myr is the most representative of our observed SEDs.
%This indicates that a single stellar component of age around 1\,Gyr in combination with a secondary component of age around 10\,Myr is the most representative of our observed SEDs.
This indicates that a post-starburst component in combination with a secondary young stellar component with ongoing SF is the most representative of our observed SEDs.

\section{Discussion and Conclusion}%\label{s:discussion}
We present a sizable sample of \SIII\ emitters identified using deep $H_2S1$ combined with deep $K_\mathrm{s}$ imaging data of ECDFS.
The deep optical and NIR imaging data in the ECDFS are utilized to construct SEDs and then derive photometric redshift, extinction and stellar mass.
Total 14 \SIII\ emitters at $z$= 1.34 and $z$= 1.23 are identified with the observed \SIII\ line flux $>2.9 \times$ 10$^{-17}$ erg s$^{-1}$ cm$^{-1}$ Hz$^{-1}$.
The rest-frame $U-V$ versus $V-J$ diagram shows that all of our \SIII\ emitters are SFGs.
None of our sample galaxies is detected in the Chandra 4\,Ms observation, although the possibility to host a Compton-thick AGN individually-undetected in the current X-ray observation can not be excluded \cite{Tan12}.
We note that the Compton-thick AGN has little influence in exciting \SIII\ lines.
We classify the morphologies of our sample based on the \textit{HST}/ACS F606W images, finding that most \SIII\ emitters are mergers and disk-type galaxies.
The available rest-frame optical morphologies from CANDELS confirm that the majority of classification while also revealing that the rest-frame UV morphologies of our sample are seriously affected by heavily dust extinction, particularly for objects which appear diffuse morphologies in rest-frame UV images.
Although the rest-frame optical lights are less affected by dust obscuration, only half of \SIII\ emission-line galaxies are covered by CANDELS.  
%Despite this, we still can see that nearly half of our \SIII\ emission-line galaxies are mergers and at least one third are disk galaxies.
The dust extinction of our sample galaxies is estimated from the modeled SEDs.
Our estimation is limited by $A_\mathrm{V}=2.75$ adopted in the sixth template (refer to Table 4) and the SED-derived extinction tends to be lower than the ture case represented by the IR to UV luminosity ration L$_\mathrm{IR}$/L$_\mathrm{UV}$.
Unlike other emission-line selected samples at similar redshift, the estimated extinction of \SIII\ emission-line galaxies appears to have no obvious correlation with luminosity, stellar mass or line equivalent width \cite{Sobral09,Dominguez13}.
The stellar masses of our sample galaxies are $8.7<\log (M/\Msun)<9.9$. 

The stellar population modeling shows that the Balmer break and the strong emission lines are the prominent features of SEDs for nearly all of the \SIII\ emitters, indicating the presence of a post-starburst and an ongoing burst components \cite{Shapley11}.
Thus, the photoionization caused by young O and B stars in ongoing burst and the collisional ionization because of supernova-driven shocks in post-starburst may all responsible for exciting strong \SIII\ lines.
This conclusion is consistent with the previous studies of \SIII\ emission lines in starburst galaxies \cite{Reines09,Rodriguez09}.

Information of some important emission lines, such as \NII, \SIII\ in the redshift range, 1.0 $<z<$ 1.5, is critical to our understanding of the history of chemical evolution.
%A better understanding of some important emission lines, such as \NII, \SIII\ in the redshift range, 1.0 $<z<$ 1.5, is critical to the buildup of the history of chemical evolution.
%The redshift range, 1.0 $<z<$ 1.5, is a significant epoch for the study of the history of chemical evolution.
%Therefore, the informations of some important emission lines, such as \NII, \SIII, at this redshift range are critical.
The on-going James Webb Space Telescope, K-band Multi-Object Spectrograph on Very Large Telescope will provide important information of optical as well as NIR emission lines.
Fundamental properties such as element abundance, metallicity and ionization state of galaxies at this redshift range can be studied and these properties will also aid researchers to reveal the buildup of heavy elements as galaxies are assembled over cosmic time and understand the cosmic reionization.

\Acknowledgements{\bahao This research uses data obtained through the Telescope Access Program (TAP), which is funded by the National Astronomical Observatories and the Special Fund for Astronomy from the Ministry of Finance. 
This work is supported by National Basic Research Program of China (973 Program 2013CB834900) and the National Natural Science Foundation of China under No. 11063002.}


\normalsize \vskip0.3in\parskip=0mm \baselineskip 18pt
\renewcommand{\baselinestretch}{1.1}\footnotesize\parindent=4mm\bahao

%\vskip0.1in \noindent %{\normalsize \bf References}
%\vskip0.1in\parskip=0mm

\end{multicols}


\begin{thebibliography}{99}
\bibitem[1]{Madau96} Madau, P., Ferguson, H.~C., Dickinson, M.~E., et al. High-redshift galaxies in the Hubble Deep Field: colour selection and star formation history to z\~{}4. Mon Not Roy Astron Soc, 1996, 283: 1388-1404

\bibitem[2]{Hopkins06} Hopkins, A.~M., Beacom, J.~F. On the Normalization of the Cosmic Star Formation History. Astron J, 2006, 651: 142-154 

\bibitem[3]{Sobral13} Sobral, D., Smail, I., Best, P.~N., et al. A large H{$\alpha$} survey at z = 2.23, 1.47, 0.84 and 0.40: the 11 Gyr evolution of star-forming galaxies from HiZELS. Mon Not Roy Astron Soc, 2013, 428: 1128-1146

\bibitem[4]{Marchesini12} Marchesini, D., Stefanon, M., Brammer, G.~B., et al. The Evolution of the Rest-frame V-band Luminosity Function from z = 4: A Constant Faint-end Slope over the Last 12 Gyr of Cosmic History. Astron J, 2012, 748: 126 

\bibitem[5]{Karim11} Karim, A., Schinnerer, E., Martínez-Sansigre, A., et al. The Star Formation History of Mass-selected Galaxies in the COSMOS Field. Astron J, 2011, 730: 61

\bibitem[6]{Sobral09} Sobral, D., Best, P.~N., Matsuda, Y., et al. Star formation at z=1.47 from HiZELS: an \Ha+\OII\ double-blind study. Mon Not Roy Astron Soc, 2012, 420: 1926-1945

\bibitem[7]{Yuan13} Yuan, T.-T., Kewley, L.~J., Richard, J. The Metallicity Evolution of Star-forming Galaxies from Redshift 0 to 3: Combining Magnitude-limited Survey with Gravitational Lensing. Astron J, 2013, 763: 9

\bibitem[8]{Yabe12} Yabe, K., Ohta, K., Iwamuro, F., et al. NIR Spectroscopy of Star-Forming Galaxies at z $\sim$ 1.4 with Subaru/FMOS: The Mass-Metallicity Relation. Publ Astron Soc Jpn, 2012, 64: 60 

\bibitem[9]{Dominguez13} Dom{\'{\i}}nguez, A., Siana, B., Henry, A.L., et al. Dust Extinction from Balmer Decrements of Star-forming Galaxies at 0.75 $<z<$ 1.5 with Hubble Space Telescope/Wide-Field-Camera 3 Spectroscopy from the WFC3 Infrared Spectroscopic Parallel Survey. Astron J, 2013, 763: 145

\bibitem[10]{Zastrow11} Zastrow, J., Oey, M.~S., Veilleux, S., et al.\  An Ionization Cone in the Dwarf Starburst Galaxy NGC 5253. Astron J, 2011, 741: L17 

\bibitem[11]{Garnett89} Garnett, D.~R. The abundance of sulfur in extragalactic H II regions. Astron J, 1989, 345: 282-297

\bibitem[12]{Perez03} P{\'e}rez-Montero, E., D{\'{\i}}az, A.~I. Line temperatures and elemental abundances in H II galaxies. Mon Not Roy Astron Soc, 2003, 346: 105-118

\bibitem[13]{Kehrig06} Kehrig, C., V{\'{\i}}lchez, J.~M., Telles, E., et al. A spectroscopic study of the near-IR \SIII\ lines in a sample of HII galaxies: chemical abundances. Astron Astrophys, 2006, 457: 477-484 

\bibitem[14]{Matrozis13} Matrozis, E., Ryde, N., Dupree, A. K. On the Galactic chemical evolution of sulphur-Sulphur abundances from the \SI$\lambda$1082 nm line in giants. Astron Astrophys, 2013, in press, arXiv:1309.0114v1

\bibitem[15]{Hagele06} H{\"a}gele, G.~F., P{\'e}rez-Montero, E., D{\'{\i}}az, et al. \ The temperature and ionization structure of the emitting gas in HII galaxies: implications for the accuracy of abundance determinations. Mon Not Roy Astron Soc, 2006, 372: 293-312

\bibitem[16]{Kistiakowsky93} Kistiakowsky, V., Helfand, D.~J. Observations of (S III) emission from Galactic radio sources - The detection of distant planetary nebulae and a search for supernova remnant emission. Astron J, 1993, 105: 2199-2210

\bibitem[17]{Diaz85a} Diaz, A.~I., Pagel, B.~E.~J., Wilson, I.~R.~G. The intensities of the sulphur III lines and the ionization mechanisms in Liners. Mon Not Roy Astron Soc, 1985, 212: 737-749

\bibitem[18]{Dopita77} Dopita, M.~A. Optical Emission from Shock Waves. II. Diagnostic Diagrams. Astrophys J Suppl Ser, 1977, 33: 437

\bibitem[19]{Binette85} Binette, L., Dopita, M.~A., Tuohy, I.~R. Radiative shock-wave theory. II - High-velocity shocks and thermal instabilities. Astron J, 1985, 297: 476-491

\bibitem[20]{Diaz85b} Diaz, A.~I., Terlevich, E., Pagel, B.~E.~J. Sulphur III lines and the excitation mechanism in NGC 1052. Mon Not Roy Astron Soc, 1985, 214: 41P-45P

\bibitem[21]{Hill99} Hill, T.~L., Heisler, C.~A., Sutherland, R., et al. Starburst or Seyfert? Using Near-Infrared Spectroscopy to Measure the Activity in Composite Galaxies. Astron J, 1999, 117: 111-125

\bibitem[22]{Ly12} Ly, C., Malkan, M.~A., Kashikawa, N., et al. The Stellar Population and Star Formation Rates of z $\sim$ 1.5-1.6 \OII-emitting Galaxies Selected from Narrowband Emission-line Surveys. Astron J, 2012, 757: 63

\bibitem[23]{Tadaki12} Tadaki, K.-i., Kodama, T., Ota, K., et al.  A large-scale structure traced by [O II] emitters hosting a distant cluster at z= 1.62.\ Mon Not Roy Astron Soc, 2012, 423: 2617-2626

\bibitem[24]{Grutzbauch12} Gr{\"u}tzbauch, R., Bauer, A.~E., J{\o}rgensen, I., et al. Suppression of star formation in the central 200 kpc of a z= 1.4 galaxy cluster. Mon Not Roy Astron Soc, 2012, 423: 3652-3662

\bibitem[25]{Kroupa01} Kroupa, P. On the variation of the initial mass function. Mon Not Roy Astron Soc, 2001, 322: 231-246 

\bibitem[26]{Oke74} Oke, J.~B. Absolute Spectral Energy Distributions for White Dwarfs. Astrophys J Suppl Ser, 1974, 27: 21 

\bibitem[27]{Puget04} Puget, P., Stadler, E., Doyon, R., et al. WIRCam: the infrared wide-field camera for the Canada-France-Hawaii Telescope. Society of Photo-Optical Instrumentation Engineers (SPIE) Conference Series, 2004, 5492: 978-987

\bibitem[28]{Wang10} Wang, W.-H., Cowie, L.~L., Barger, A.~J., et al. Ultradeep K$_{S}$ Imaging in the GOODS-N. Astrophys J Suppl Ser, 2010, 187: 251-271

\bibitem[29]{Hsieh12} Hsieh, B.-C., Wang, W.-H., Hsieh, C.-C., et al. The Taiwan ECDFS Near-Infrared Survey: Ultra-deep J and K$_{S}$ Imaging in the Extended Chandra Deep Field-South. Astrophys J Suppl Ser, 2012, 203: 23 

\bibitem[30]{Gawiser06} Gawiser, E., van Dokkum, P.G., Herrera, D., et al. The Multiwavelength Survey by Yale-Chile (MUSYC): Survey Design and Deep Public UBVRIz' Images and Catalogs of the Extended Hubble Deep Field-South. Astrophys J Suppl Ser, 2006, 162: 1-19

\bibitem[31]{Rix04} Rix, H.-W., Barden, M., Beckwith, S.V.W., et al. GEMS: Galaxy Evolution from Morphologies and SEDs. Astrophys J Suppl Ser, 2004, 152: 163-173

\bibitem[32]{Caldwell08} Caldwell, J.~A.~R., Kocevski, D.D., Faber, S.M., et al. GEMS Survey Data and Catalog. Astrophys J Suppl Ser, 2008, 174: 136-144

\bibitem[33]{Grogin11} Grogin, N.~A., Kocevski, D.D., Faber, S.M., et al. CANDELS: The Cosmic Assembly Near-infrared Deep Extragalactic Legacy Survey. Astrophys J Suppl Ser, 2011, 197: 35

\bibitem[34]{Koekemoer11} Koekemoer, A.~M.,Faber, S. M.; Ferguson, Henry C., et al. CANDELS: The Cosmic Assembly Near-infrared Deep Extragalactic Legacy Survey-The Hubble Space Telescope Observations, Imaging Data Products, and Mosaics.\ Astrophys J Suppl Ser, 2011, 197: 36

\bibitem[35]{Bertin96} Bertin, E., Arnouts, S. SExtractor: Software for source extraction. Astron Astrophys Suppl Ser, 1996, 117: 393-404

\bibitem[36]{Bunker95} Bunker, A.~J., Warren, S.~J., Hewett, P.~C., et al. On near-infrared \Ha\ searches for high-redshift galaxies. Mon Not Roy Astron Soc, 1995, 273: 513-516 

\bibitem[37]{Geach08} Geach, J.~E., Smail, I., Best, P.~N., et al. HiZELS: a high-redshift survey of H{$\alpha$} emitters - I. The cosmic star formation rate and clustering at z = 2.23. Mon Not Roy Astron Soc, 2008, 388: 1473-1486

\bibitem[38]{Cardamone10} Cardamone, C.~N., van Dokkum, P.G., Urry, C.M., et al. The Multiwavelength Survey by Yale-Chile (MUSYC): Deep Medium-band Optical Imaging and High-quality 32-band Photometric Redshifts in the ECDF-S. Astrophys J Suppl Ser, 2010, 189: 270-285 

\bibitem[39]{Brammer11} Brammer, G.~B., van Dokkum, P.~G., Coppi, P. EAZY: A Fast, Public Photometric Redshift Code.\ Astron J, 2008, 686: 1503-1513 

\bibitem[40]{Lee12} Lee, J.~C., Ly, C., Spitler, L., et al. A Dual-Narrowband Survey for \Ha\ Emitters at Redshift of 2.2: Demonstration of the Technique and Constraints on the \Ha\ Luminosity Function. Publ Astron Soc Pacific, 2012, 124: 782-797

\bibitem[41]{Hayes10} Hayes, M., Schaerer, D., {\"O}stlin, G. The H-alpha luminosity function at redshift 2.2 . A new determination using VLT/HAWK-I. Astron Astrophys, 2010, 509: L5

\bibitem[42]{Fioc97} Fioc, M., Rocca-Volmerange, B. PEGASE: a UV to NIR spectral evolution model of galaxies. Application to the calibration of bright galaxy counts. Astron Astrophys, 1997, 326: 950-962

\bibitem[43]{Shapley11} Shapley, A.~E. Physical Properties of Galaxies from z = 2-4. Ann Rev Astron Astrophys, 2011, 49, 525-580 

\bibitem[44]{Williams09} Williams, R.~J., Quadri, R.~F., Franx, M., et al. Detection of Quiescent Galaxies in a Bicolor Sequence from Z = 0-2. Astron J, 2009, 691: 1879-1895

\bibitem[45]{Wuyts08} Wuyts, S., Labb{\'e}, I., Schreiber, N.M.F., et al. FIREWORKS U$_{38}$-to-24 {$\mu$}m Photometry of the GOODS Chandra Deep Field-South: Multiwavelength Catalog and Total Infrared Properties of Distant K$_{s}$-selected Galaxies. Astron J, 2008, 682: 985-1003 

\bibitem[46]{Xue11} Xue, Y.~Q., and 24 colleagues. The Chandra Deep Field-South Survey: 4 Ms Source Catalogs.\ Astrophys J Suppl Ser,  2011, 195: 10

\bibitem[47]{Lotz04} Lotz, J.~M., Primack, J., Madau, P. A New Nonparametric Approach to Galaxy Morphological Classification. Astron J, 2004, 128: 163-182 

\bibitem[48]{Stott13} Stott, J.~P., Sobral, D., Smail, I., et al. The merger rates and sizes of galaxies across the peak epoch of star formation from the HiZELS survey. Mon Not Roy Astron Soc, 2013, 430: 1158-1170 

\bibitem[49]{Kriek09} Kriek, M., van Dokkum, P.~G., Labb{\'e}, I., et al. An ultra-deep Near-infrared spectrum of a compact quiescent galaxy at $z$ =2.2. Astron J, 2009, 700: 221

\bibitem[50]{Maraston05} Maraston, C. Evolutionary population synthesis: models, analysis of the ingredients and application to high-z galaxies. Mon Not Roy Astron Soc, 2005, 362, 799-825

\bibitem[51]{Wuyts11} Wuyts, S., F{\"o}rster Schreiber, N.M., van der Wel, A., et al. Galaxy Structure and Mode of Star Formation in the SFR-Mass Plane from z $\sim$ 2.5 to z $\sim$ 0.1. Astron J, 2011. 742: 96

\bibitem[52]{Calzetti00} Calzetti, D., Armus, L., Bohlin, R.~C., et al. The Dust Content and Opacity of Actively Star-forming Galaxies. Astron J, 2000, 533, 682-695 

\bibitem[53]{Kennicutt12} Kennicutt, R.~C., Evans, N.~J. Star Formation in the Milky Way and Nearby Galaxies.\ Ann Rev Astron Astrophys, 2012, 50: 531-608

\bibitem[54]{Tan12} Tan Y, Wang J X, Zhang K. On the nature of X-ray “unobscured” Seyfert 2 galaxies. Sci China-Phys Mech Astron, 2012, 55: 2482-2491

\bibitem[55]{Reines09} Reines, A.~E., Johnson, K.~E., Hunt, L.~K. A New View of the Super Star Clusters in the Low-Metallicity Galaxy SBS 0335-052. American Astronomical Society Meeting Abstracts, 2009, \#214 214, \#418.01

\bibitem[56]{Rodriguez09} Rodr{\'{\i}}guez-Ardila, A., Riffel, R., Pastoriza, M.~G., et al. Detection of Hidden Starburst in Active Galactic Nuclei by Means of Near-Infrared Spectroscopy. The Starburst-AGN Connect, 2009, 408: 203 

\end{thebibliography}
\end{document}